\documentclass[12pt,preprint]{aastex}

\usepackage{epsfig}

\newcommand{\etal}{et al.~}

\shorttitle{Study Accretion Disk \& Corona With Simulation}
\shortauthors{Yao \etal}

\begin{document}


\title{Studying the Properties of Accretion Disks and Coronae in
Black Hole X-ray Binaries with Monte-Carlo Simulation
}


\author{Yangsen Yao\altaffilmark{1,2,3}, S. Nan Zhang\altaffilmark{1,2},
Xiaoling Zhang\altaffilmark{1,2}, Yuxin Feng\altaffilmark{1, 2},
Craig R. Robinson\altaffilmark{4}}

\altaffiltext{1}{Physics Department, University of Alabama in Huntsville,
Huntsville, AL 35899; yaoy@email.uah.edu, zhangsn@email.uah.edu,
zhangx@email.uah.edu, fengyx@jet.uah.edu}
\altaffiltext{2}{National Space Science and Technology Center,
320 Sparkman DR., SD50, Huntsville, AL 35805}
\altaffiltext{3}{Present address: Department of Astronomy, University of
Massachusetts, Amherst, MA 01003; yaoys@astro.umass.edu}
\altaffiltext{4}{Division of Information Systems, National Science Foundation,
4201 Wilson Blvd, Arlington, VA 22230;crobinso@nsf.gov}

\begin{abstract}
Understanding the properties of the hot corona is important for
studying the accretion disks in black hole X-ray binary systems. Using the
Monte-Carlo technique to simulate the inverse Compton
scattering between photons emitted from the cold disk and electrons
in the hot corona,
we have produced two table models in the $XSPEC$ format for
the spherical corona case and the disk-like (slab) corona case.
All parameters in our table models are physical properties of the
system and can be derived from data fitting directly.
Applying the models to broad-band spectra of the black hole
candidate XTE J2012+381 observed with BeppoSAX,
we estimated the size of the corona and the inner radius of the disk.
The size of the corona in this system
is several tens of gravitational radius,
and the substantial increase of the inner disk radius
during the transit from hard-state to soft-state is
not found.
\end{abstract}


\section{Introduction}
A black hole X-ray binary (BHXB) system consists of a black hole
(a compact object with mass $>3M_{\odot}$) and its companion star.
Material can be transferred from the companion to the black hole
either through Roche lobe overflow or via a stellar wind
(accretion process). Because the accreted matter must lose its
angular momentum before it can be ``swallowed'' by the black hole,
an accretion disk is usually formed and the gravitational
potential energy of the accreted matter is released through
radiation. The X-ray spectrum of a black hole X-ray binary system
usually can be well fit with a two-component model: a
black-body-like component and a power-law-like component (Tanaka
\& Lewin 1995). The black-body-like component turns off above 20
keV and is believed to be emitted from the accretion disk. The
power-law-like component can extend up to several hundred keV,
which suggests that a high temperature electron cloud (corona)
exists above the accretion disk and the inverse Compton scattering
between the disk photons and electrons in the corona is the main
mechanism to produce this component.

In the X-ray astronomy community, the multi-color disk model (MCD)
({\em diskbb} in {\em XSPEC}, Mitsuda \etal 1984; Makishima \etal 1986)
plus a power-law (PL) model has been employed traditionally to fit the energy
spectra of BHXBs, then the parameters of the MCD
model are used to infer the physical parameters of the accretion system.
However, this model is over-simplified in the following two aspects.

First, the assumption that the power-law component extends
straight to the low-energy limit of the spectrum is unreasonable,
because the power-law component is believed to be produced by
inverse Compton scattering. The most natural source of the seed
photons is the thermal radiation from the accretion disk; they are
of the same origin as the black-body-like component. Therefore a
low-energy cutoff must be present in the power-law
component, because the seed photon distribution has a peak energy,
below which there are only a small amount of seed photons. Neglecting this
low-energy cutoff by applying a simple power-law spectrum model
would under-estimate the flux in the black-body-like component.

Second, it has been a rather standard approach by many (including
some authors of this paper) in the field to take the
black-body-like flux derived from the spectral model fitting as
the true flux of the accretion disk and infer the inner disk
radius from it. This ignores the radiative transfer process in the
production of the power-law component. Since the power-law
component is likely produced by scattering the soft photons in the
original black-body-like component from the accretion disk, each
photon in the power-law component comes at the expense of a lost
photon in the black-body-like component, even if no absorption
occurs in the Comptonization process. Therefore certain
corrections are needed to infer the original flux of the
black-body-like component, otherwise the real flux would be
under-estimated, as realized by some authors (e.g., Kubota,
Makishima \& Ebisawa 2001), which might be the reason that the
inner disk radius varies significantly when a BHXB
transits from its hard state to its soft state or vice versa, like
in XTE J1550-564 (Sobczak \etal 1999a), GRO J1655-40 (Sobczak
\etal 1999b), XTE J2012+381 (Campana \etal 2002), etc..

In order to estimate the true flux of the accretion disk and then
to estimate the inner disk radius, one needs to establish an
intrinsic relation between the soft (low energy) photons and the
hard (high energy) photons, then to make a radiative transfer
correction to recover the original flux of the accretion disk,
which makes it inevitable to calculate the Comptonized spectrum.
The Comptonized spectrum has been computed by many authors using
different approaches.

One of the analytical approaches was
carried out by
assuming an optically thick non-relativistic plasma and then solving the
Kompaneets equation either numerically or analytically (e.g., Sunyaev \&
Titarchuk 1980; Payne 1980). This approach was improved later
by Titarchuk (1994, T94 hereafter) and Titarchuk \& Lyubarskij
(1995) so that it could be applied to both non-relativistic and
relativistic plasma, and  was further verified by Hua \& Titarchuk
(1995, hereafter HT95) with
different analytical approximations and Monte-Carlo simulations.

Another approach is to calculate radiative transfer with
Monte-Carlo simulation (e.g., Corman 1970; Loh \& Garmire 1971;
Pozdnyakov, Sobol' \& Sunyaev 1977; Fenimore \etal 1982; etc.).
This method has been implemented and described in detail by
Pozdnyakov, Sobol' \&
Sunyaev (1977, PSS77 hereafter), under the assumption
that the probability for a
collision between a photon and an electron is independent of the
electron energy and the scattering angle.
It was then significantly
improved by Gorecki \& Wilczewski (1984) by removing the above
inaccurate assumption.
The works mentioned
above take the Compton scattering process as the main process when
computing the Comptonized spectra from hot plasma, some other
authors also take into account other important physical processes
such as
bremsstrahlung, pair production (e.g., Skibo \etal 1995) and
disk-reflection (e.g., Haardt \& Maraschi 1991).

Recently, Poutanen \& Svensson (1996, hereafter PS96), by taking into account
almost all important physical processes in the hot plasma and solving
the radiative transfer equation iteratively for each scattering,
built a new model to describe the emergent spectra from the plasma, which has
been successfully tested with Monte-Carlo simulations (Stern \etal 1995).

The above works have been applied to both AGNs and X-ray binaries
(e.g., Iwasawa \etal 2004; Wardzinski \etal 2002; Petrucci \etal 2000;
Miller \etal 2004).
In this paper, we only consider the Compton scattering process
in the hot plasma
(corona) and use the Monte-Carlo simulation to compute the
emergent photon spectra. We stress on building a simple, self-consistent model
in which the soft photons and the hard photons can be intrinsically linked
and the radiative transfer process can be taken into account automatically.
Therefore, the physical parameters of the corona, as well as the parameters
of the accretion disk, can be directly derived from the spectral fitting with
the model. We will discuss the difference between our model and the power-law
model, especially in deriving the inner disk radius when applied to BHXBs.

This paper is organized as following. In section 2,
our Monte-Carlo simulation is introduced, in section 3 the
results of applying
 our table models to the source XTE J2012+381 is reported, and summary
and discussion are presented in section 4.

\section{Monte-Carlo Simulations and Table Models}
\subsection{Assumptions and Simulations}
In our simulations, we assume that the accretion disk is an
optically thick Keplerian disk (Mitsuda \etal 1984), and during
the accretion process, the accretion rate ($\dot{m}$) is constant
and the gravitational potential energy loss of the accreted
material is radiated away in black-body radiation locally.
Therefore the temperature radial profile is $T(r) \sim r ^{-3/4}$,
as in the MCD model (Mitsuda \etal 1984; Makishima \etal 1986). We
also assume that the corona above the accretion disk may take
either spherical or disk geometry, the electron density
distribution in the corona is uniform, and the electron energy
distribution in the corona follows a thermal form as assumed in previous
works (HT95, T94, PS96); the disk reflection is ignored,
and the corona is in stationary equilibrium.

We sample photons for a given combination of the parameters,
i.e., the temperature at the inner boundary of the
accretion disk, the size of the corona (for a spherical corona
system, the size is defined as the value of its radius; for a
disk-like corona system, it is defined as the vertical height),
the temperature of the electrons in the corona, and the optical depth of
the corona (defined as $\tau = \int n_e \sigma_T dl$, where $n_e$
is electron density, $\sigma_T$ is Thompson scattering cross
section and the integral is along the radial direction in the
spherical corona case or along the vertical direction in
disk-like corona case).
Since the radiation is taken as a black body locally,
the number
of photons released from a ring on the accretion disk is,
\begin{equation} \begin{array}{rll}
dN_{ph} &=& a^*T^32\pi r dr\\
    &\propto& r^{-5/4} dr,
\end{array}
\end{equation}
where
\[ a^*= \frac{2\pi}{c^2}\left(\frac{k}{h}\right)^3 \int^\infty_0 \frac{x^2dx}{e^x -1}, \]
$c$ is the speed of light, $k$ is Boltzmann's constant and
$h$ is Planck's constant.
The photon energies follow blackbody distribution with temperature $T(r)$ and
the initial direction of a photon is sampled uniformly in $4\pi$ solid angles.
In a spherical corona system, the corona covers a portion of the disk,
then a photon emitted from the disk
may or may not enter the corona,
depending upon the
emitting location and the initial direction.
In disk-like corona systems the corona covers the whole disk,
so a photon emitted from the disk always enters the corona first.

After a photon enters the hot corona, the probability ($\rho$) for it
to interact with an electron is
determined by
its free-path distance,
\begin{equation}
d\rho  \propto  e ^ {- \tau_l} d\tau_l,
\end{equation}
and
\begin{equation}
d\tau_l  =  \sigma_{K-N} n_edl, \label{freepath}
\end{equation}
where $\tau_l$ is the optical depth of the free-path distance along the moving
direction,
$\sigma_{K-N}$ is Klein-Nishina cross section \citep{ber72},
\begin{equation}
\sigma_{K-N}(x) = 2\pi r^2_0 \frac{1}{x}\left[ \left(1 - \frac{4}{x} -
        \frac{8}{x} \right)
        \ln (1 +x) + \frac{1}{2} + \frac{8}{x} -
        \frac{1}{2(1+x)^2} \right],
\end{equation}
and
\[ x = \frac{2h \nu }{m_ec^{2}} \gamma (1 - \overline{v} \cdot
        \overline{\Omega}/c), \]
$h\nu$, $\overline{\Omega}$ are the energy and the direction of
the incident photon, respectively,
and $\overline{v}$ is the electron velocity, all in the laboratory frame;
$r_0$ is the classical electron
radius ($r_0 = e^2/mc^2$) and $m_ec^2$ is the electron rest energy.
It is obvious that $\sigma_{K-N}$ depends
on the incident photon energy, electron energy and their moving directions.
In order to determine the photon
free-path distance $l$ in equation~\ref{freepath}, we need to calculate
$\sigma_{K-N}$, therefore we need to know the electron
energy ($E_e$) along a sampled direction, which usually is a function of the
photon's free-path distance, i.e., the dependency relation for
equation~\ref{freepath} is like,
$l \rightarrow \sigma_{K-N} \rightarrow E_e \rightarrow l$. To simplify
this problem, we assume that the energy distribution in the corona is
uniform, then the dependency between $E_e$ and $l$ is no longer needed.
However, the sampling of the electron velocity is not trivial, because
it is related to the input photon information and the total Compton
cross section $\sigma_{K-N}(x)$,
\begin{equation}
\rho(\overline{v}) \propto N(\overline{v})(1-\overline{v} \cdot
   \overline{\Omega}/c)\sigma_{K-N}(x),
\label{equ:electron}
\end{equation}
where, $N(\overline{v})$ is electron distribution which is further
assumed to be isotropic and follows a thermal form $dN_e \propto
e^{-E_e / kT} \sqrt{E_e} dE_e$, $T$ is the temperature of the corona.

Equation \ref{equ:electron} is a two-dimensional probability
distribution, which is rather complicated to obtain and has been
addressed in previous works \citep{sob74,goe84}. In our
simulation, we adopt the algorithm developed by
\citet{hua97} to sample the electron energy $E_e$ and its moving
direction $\overline{\Omega}_e$.
Therefore $\sigma_{K-N}$ can be calculated (the photon energy and
moving direction are known) and the photon's free-path distance
can be derived using equation~\ref{freepath}.

After each interaction, the scattered photon direction can be
sampled by using the differential cross section \citep{akh65},
which is a function of
electron velocity ($\overline{v}$),
the incident photon energy and direction, and the scattered photon
energy. In the electron's rest frame, the formula is
relatively simple \citep{kle29},
\begin{equation} \label{differentialequation}
\frac{d\sigma}{d\Omega} = \frac{1}{2}r^2_0 \left(\frac{h\nu}{h\nu_i} \right)
      \left( \frac{h\nu}{h\nu_i} + \frac{h\nu_i}{h\nu} - \sin^2\theta \right),
\end{equation}
where $\theta$ is the scattering angle, $h\nu_i$ and $h\nu$ are respectively
the incident photon energy
and the scattered photon energy in the
electron's rest frame and related to each other by
\begin{equation} \label{scatteredphotonenergy}
h\nu = \frac{h\nu_i}{1 + \frac{h\nu_i}{m_ec^2}(1-\cos \theta )}.
\end{equation}
Therefore, after sampling the electron energy and direction, we transform
the incident photon information to the electron's rest frame, and substitute
the scattered photon energy ($h\nu$) in equation~\ref{differentialequation}
with equation~\ref{scatteredphotonenergy}, then the scattered photon
direction can be sampled. In the end, the scattered photon energy
can be calculated easily.

After zero or more scattering with electrons, an
incident photon will escape and the escaped photons will form a
spectrum, which depends on the inclination angle.
In our simulation, we ignore the disk reflection, i.e., a photon
is assumed to be absorbed by the accretion disk when it collides
with the disk.

It is worth noting that unlike the method described by Gorecki \&
Wilczewski (1984), we trace each single photon from its emitted
location until it escapes from the disk-corona system, then record
it as a single event, i.e., each recorded photon has exactly the
same statistical weight as others. In this manner, we sacrifice
the simulation efficiency to avoid calculating the escape
probability and the scattering probability distribution in each
scattering. Therefore our Monte-Carlo code has been sufficiently
simplified and its efficiency obviously depends on the
configuration of the disk-corona system (especially on the optical
depth of the corona); the total computation time is also tolerable
(e.g., it takes 5 minutes to produce 500,000 events for a
spherical corona with optical depth around unity when running on a
1.50 GHz Intel Xeon CPU).

\subsection{Simulation Summary and Comparison with the Previous Works}

\begin{figure}
\plotone{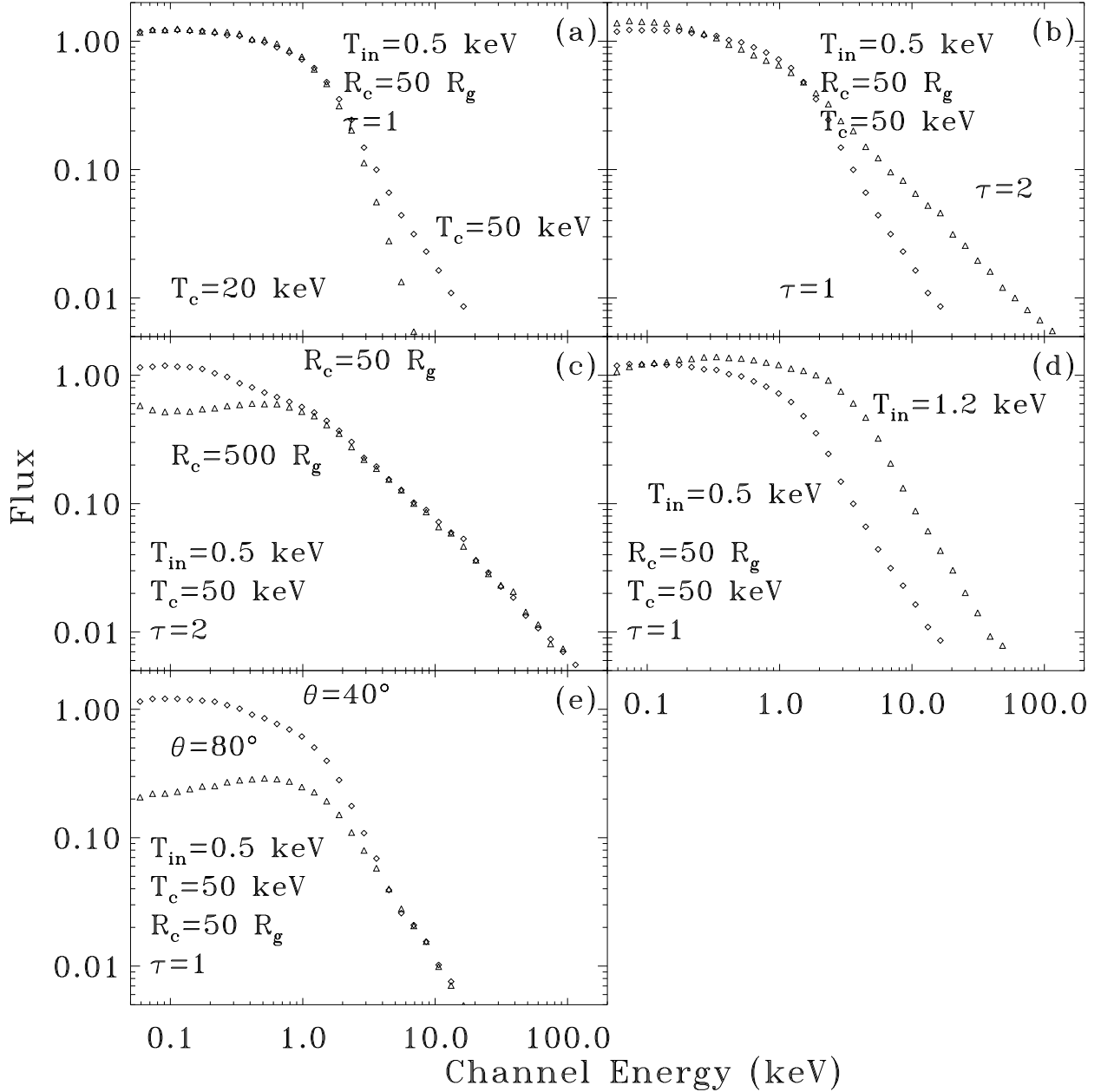} \caption{ The angle-averaged simulated spectrum
as a function
  of (a) corona temperature, (b) corona opacity, (c) corona size, (d) inner
  disk temperature, and emergent spectra as a function of the disk inclination
  angle (e).} \label{yParameter}
\end{figure}

\begin{figure}
\plotone{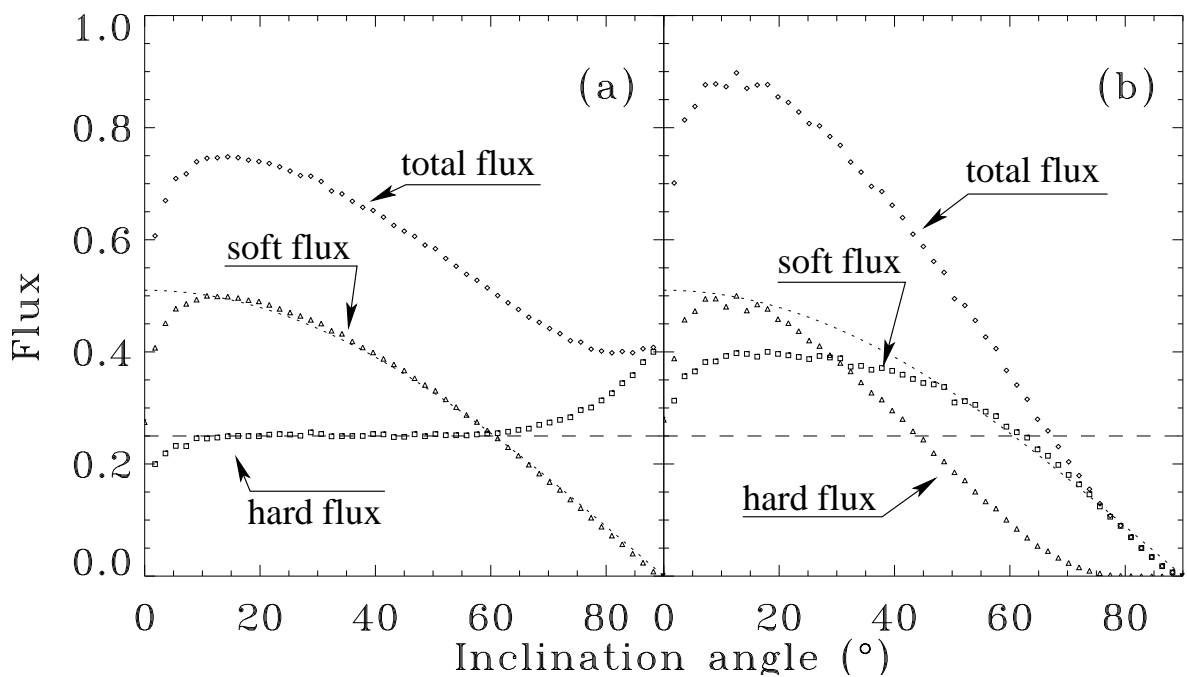} \caption{ The flux dependency on the inclination angle in a
  spherical corona system (a) and in a disk-like corona system (b).
  The {\sl dotted line} indicates a scaled cosine function and the
  {\sl dashed line} indicates a constant value. The {\sl hard flux} indicates
  the photons with at least one scattering and the {\sl soft flux} indicates
  the photons without any scattering. The turn-off in the low angle end is
  artificial (the counting statistic is poor).} \label{angularDependency}
\end{figure}

For a spherical corona system, Figure~\ref{yParameter} summarizes the
dependency of simulated
spectra on the input corona parameters (corona temperature $T_c$,
opacity $\tau$, and size $R_c$) as well as on the
input disk properties (inner disk temperature $T_{in}$ and
disk inclination angle $\theta$).
For a disk-like corona, the emergent spectra have the similar
dependency except for $R_c$ and $\theta$, because in this case,
$R_c$ is defined as the vertical height of the corona and the
corona always covers the accretion disk, therefore
$R_c$ is solely determined by $\tau$ under the uniform density assumption.
For the dependency on $\theta$, in a spherical corona system,
the flux of the soft component of the
spectrum at different inclination angle differs by a cosine factor due to
the projection effect,
whereas
the hard component is nearly isotropic except for at very high $\theta$
(Figure~\ref{angularDependency}.a) because
$\tau$ does not depend on $\theta$.
Strictly speaking, the effective optical depth is slightly different
for different $\theta$ if we consider the seed photon
distribution along the accretion disk and this difference is obvious
when the inclination angle of the disk is high. However,
in a disk-like corona system, in addition to the effective area due to the
projection, the effective optical depth of the corona also depends on the
viewing angles, therefore both the soft flux and the hard flux are related to
$\theta$ by a cosine-like factor (Figure~\ref{angularDependency}.b).

\begin{figure}
\plotone{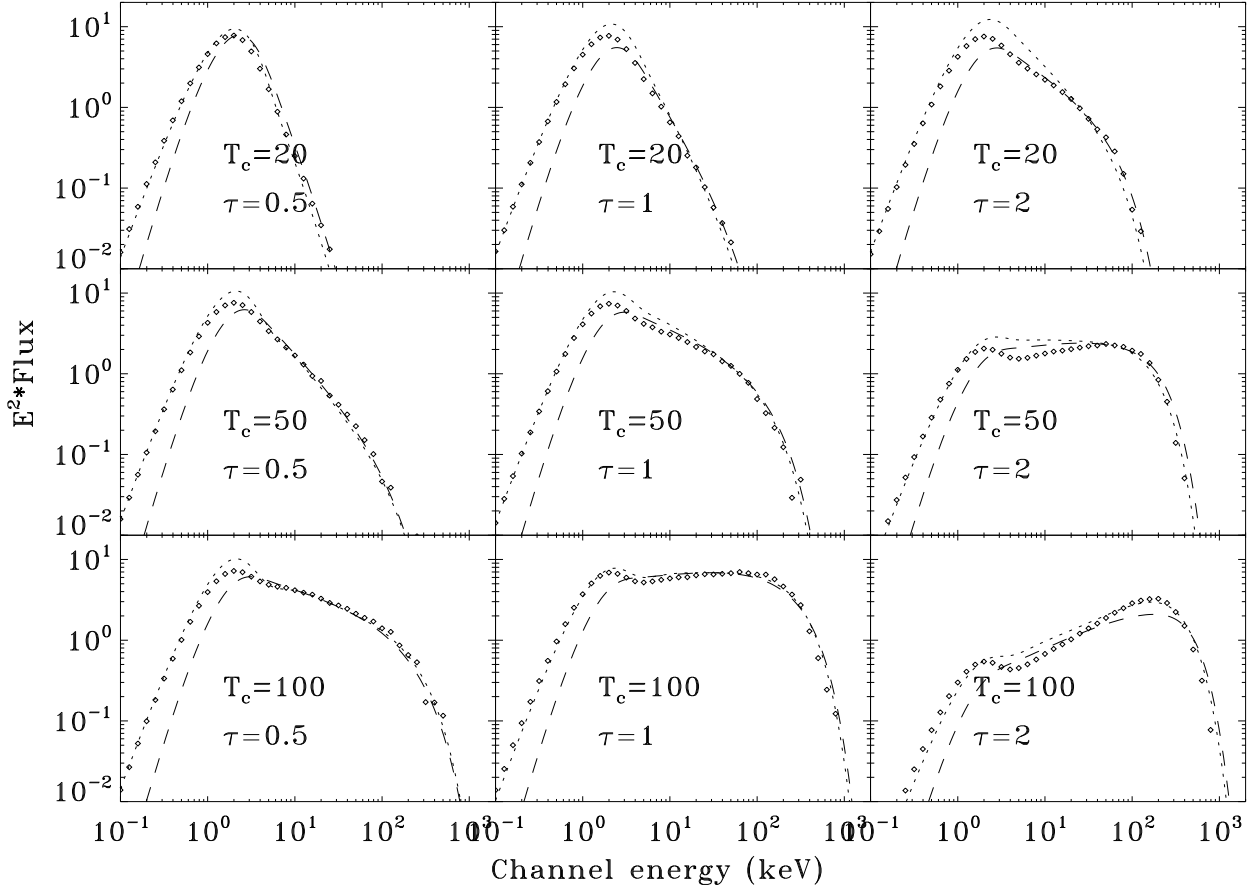} \caption{ Angle-averaged Comptonized spectra
  E$^2$*Flux from a spherical corona.
  The incident photons are assumed from the center of the corona.
  The {\sl diamond symbols}: results from our simulation; {\sl dotted
    line}: results from \citet{pou96}; {\sl dashed line}: results from
  \citet{tit94}. $T_c$ is the electron thermal temperature
  in the corona and $\tau$ is the Thomson optical depth of the
  corona. See text for detail.} \label{comparisonT94PS96}
\end{figure}

We compare the angle-averaged Comptonized spectra calculated from
our simulation with those from the analytic formulae for thermal
Comptonization by T94, and with those obtained
from the iterative scattering method by PS96 (Figure~\ref{comparisonT94PS96}).
In our simulation and the calculation by PS96, the incident photons are
from the center of a spherical corona and follow the Planck's formula,
whereas in the computation by T94, the incident photons were assumed to follow
the Wien form. The key parameters of the corona, $T_c$ and $\tau$,
are also listed in Figure~\ref{comparisonT94PS96}.
The comparison implies that, for our work and the works of T94 and PS96,
the results are fairly consistent with each other when the corona is optically
thin ($\tau_c \le 1$); when the corona is optically thick
($\tau_c \ge 2$) the differences between all the works become visible.

\subsection{Table models}
Based on the simulated spectra, we have built two $XSPEC$ format table
models\footnote{ftp://legacy.gsfc.nasa.gov/caldb/docs/memos/ogip\_92\_009/ogip\_92\_009.ps},
one for spherical corona system and one for disk-like corona system. The table models
consist of the following parameters: temperature at the inner
boundary of the accretion disk ($T_{in}$), thermal electron
temperature in the corona ($T_c$) , radial size of the spherical corona
($R_c$, in unit of $R_g=GM/c^2$, where $M$ is the black hole mass),
optical depth of the corona ($\tau$), inclination angle of
the accretion disk ($\theta$), and normalization parameter
($K_{norm}$) which is added by the $XSPEC$ automatically and
related to the disk radius by
$K_{norm}=((R_{in}/km)/(D/10kpc))^2$, where $D$ is the distance of
the source (see Appendix for detail). Because for the
disk-like corona $R_c$ is redundant, it does not appear
in the table models for the disk-like corona.
In \S3, we will apply these two models to
BHXB XTE~J2012+381.

\section{Application to XTE J2012+381}
\begin{figure}
\vbox{
  \hbox{
    \psfig{figure=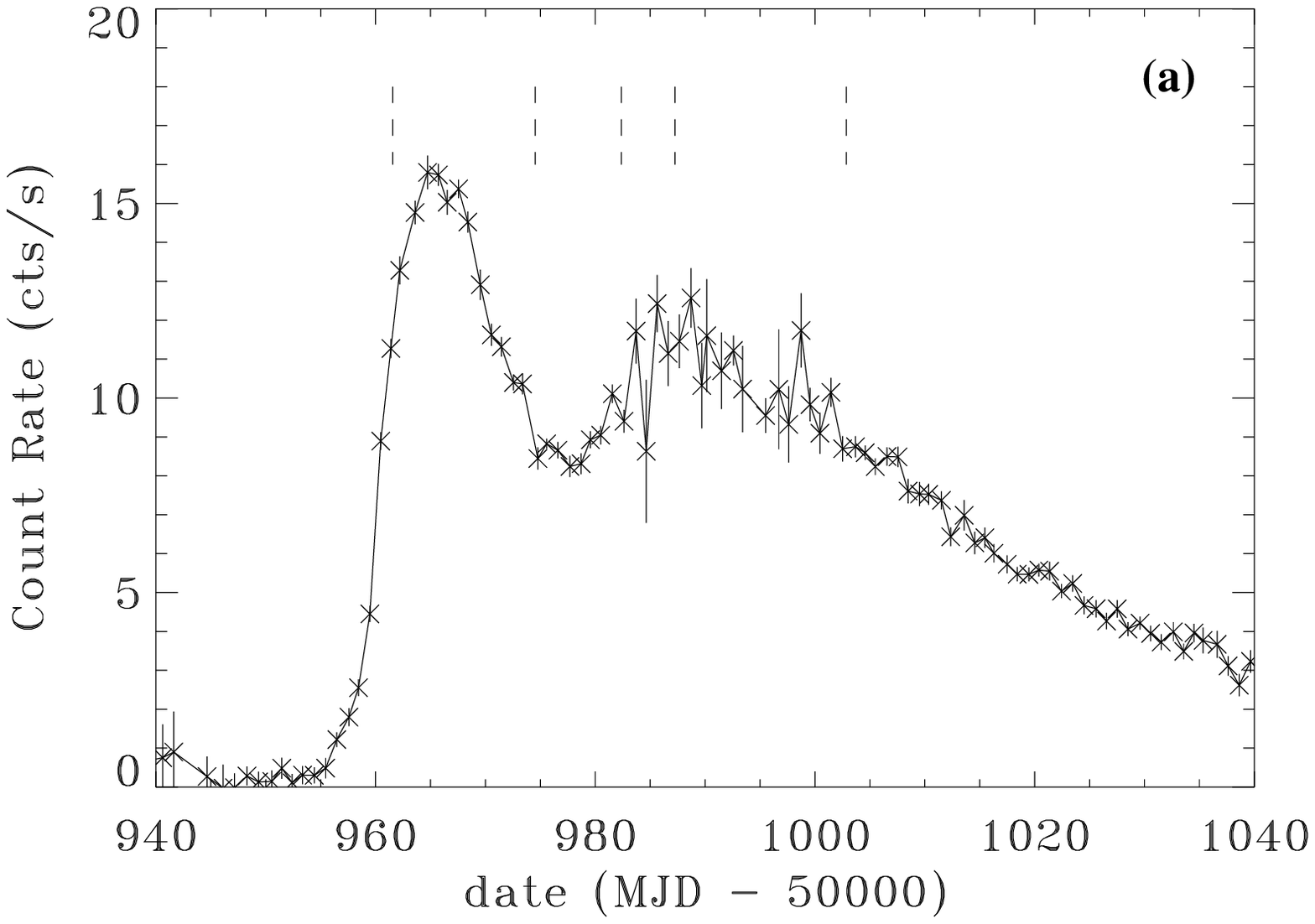,width=0.45\textwidth} \hspace{0.1in}
    \psfig{figure=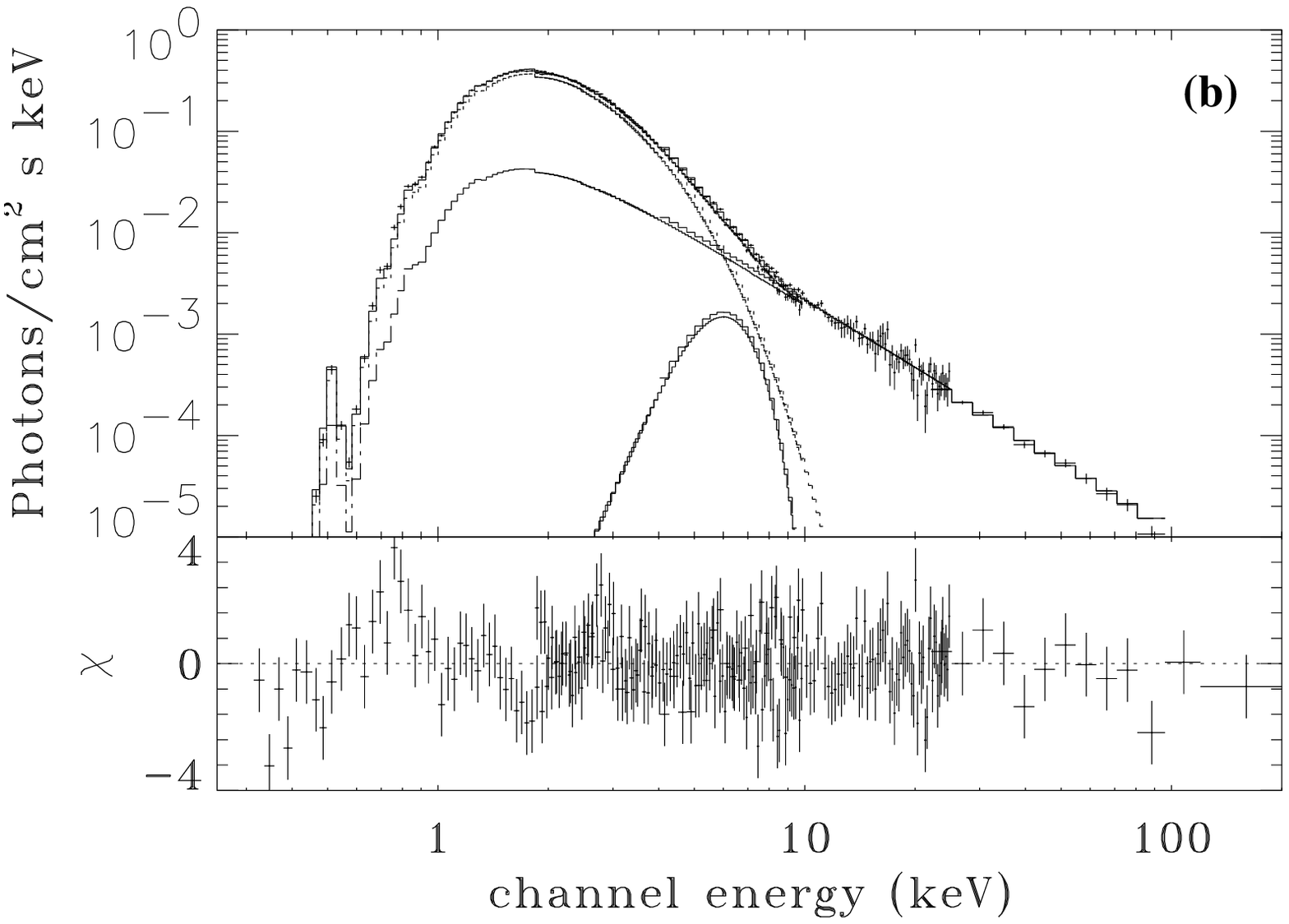,width=0.45\textwidth} }
  \hbox{
    \psfig{figure=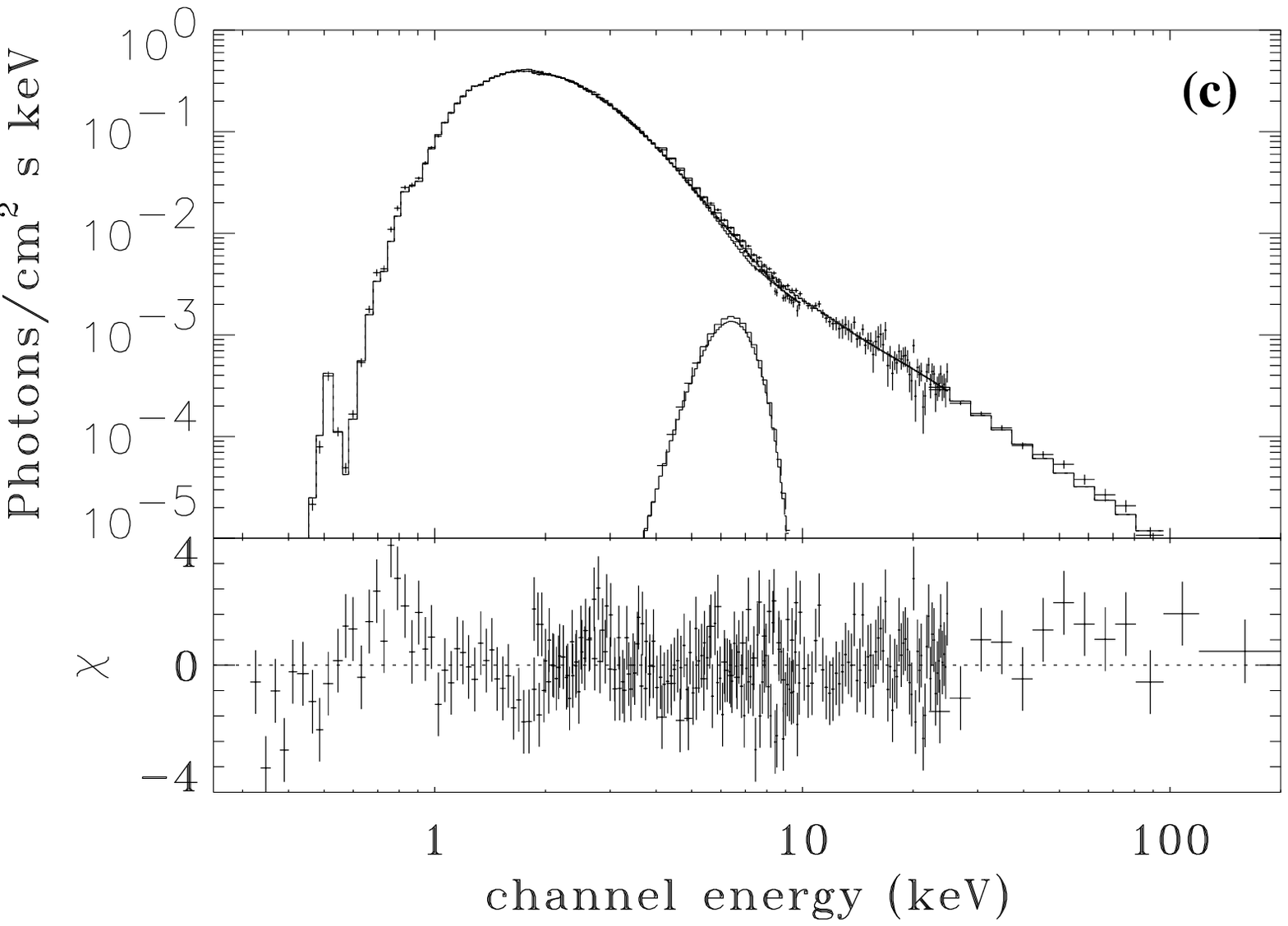,width=0.485\textwidth}
    \psfig{figure=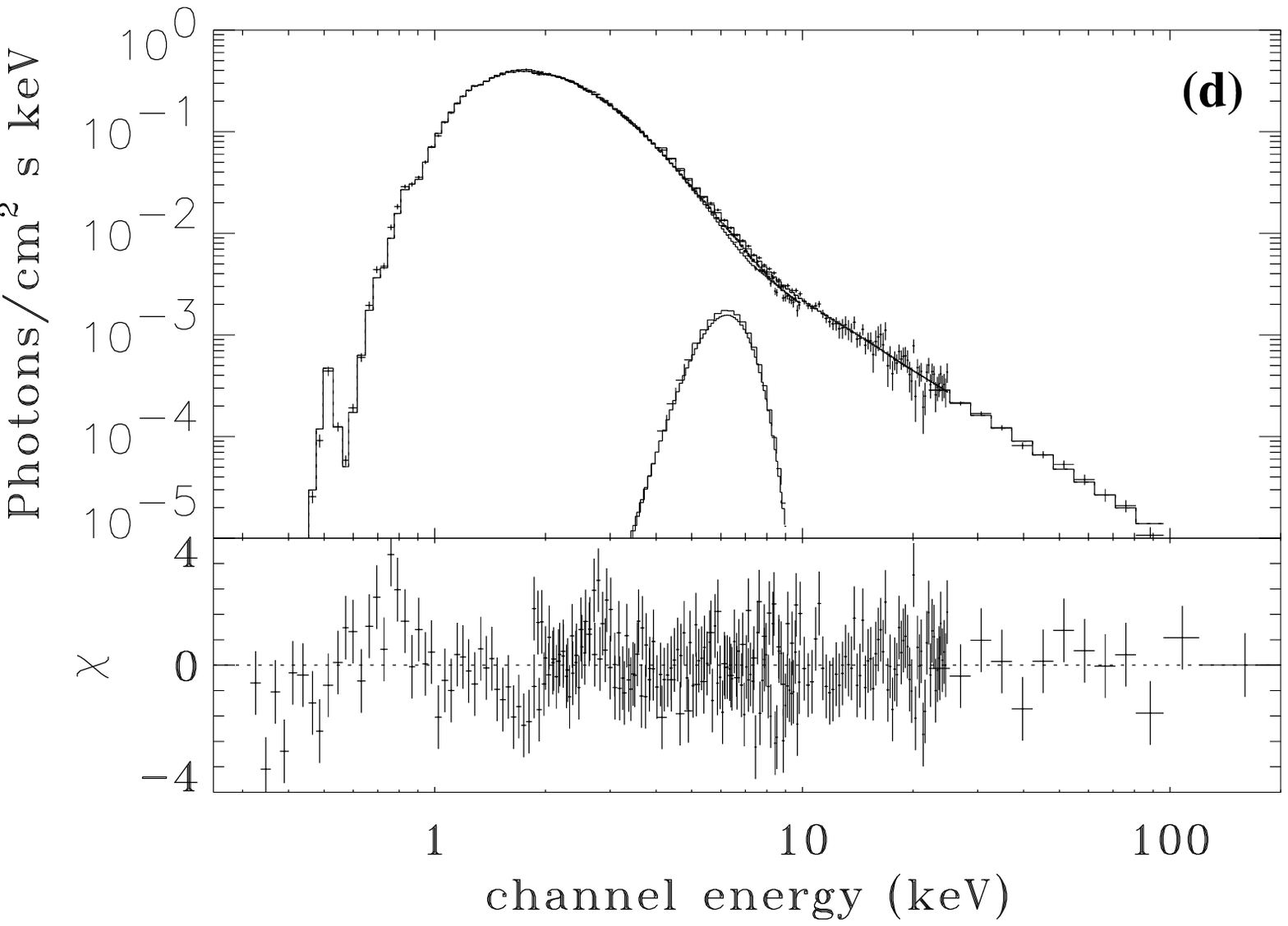,width=0.485\textwidth} }
}
\caption{ Panel (a) is the ASM/RXTE light curve
  of the black hole candidate XTE J2012+381 during its outburst in
  1998; the dotted lines indicate the dates of BeppoSAX observations
  we have analyzed. The first observation is in the rising phase in
  which the hard component is relatively strong whereas the other
  four observations are in the soft state. Panel (b) shows the
  unfolded spectrum and the residuals for the first observation
  modeling with MCD+PL with inter-stellar absorption and
  a Gaussian line. The best fit parameters are: $N_H = 1.36 \pm 0.02
  \times 10^{22}$ cm$^{-2}$, T$_{in}=0.75\pm0.01$ keV, $K_{BB} =
  1097\pm50,\ \Gamma=2.22\pm0.04, norm= 0.32\pm 0.05, E_{gauss} =
  6.1 \pm 0.2$ keV, $\sigma_{gauss} = 1.0\pm 0.2$ keV ,
  $norm_{gauss} = (3.5\pm0.4)\times 10^{-3}$, and $\chi^2$ is
  277 with 244 degrees of freedom. Panels (c) and (d) are
  unfolded spectra and the residuals for the first observation
  modeling with our table models for a spherical corona system
  and for a disk-like corona system respectively.
  The best fit parameters are presented
  in Table~\ref{modelParameters}. } \label{spectralModel}
\end{figure}

The X-ray transient XTE J2012+381
was discovered with the All Sky Monitor aboard Rossi X-ray Timing Explorer
on May 24, 1998 (Remillard \etal 1998). Its spectra observed
with ASCA consist of a soft thermal component (with temperature around 0.8
keV) and a hard power-law tail (with photon index around 3) (White \etal
1998), which is considered to be indicative of a
black hole candidate (Lewin, Van Paradijs \& Van Den Heuvel 1995).
Five observations were carried out with
BeppoSAX from 1998 May 28 to 1998 July 8 (Fig~\ref{spectralModel}.a).
The spectra with the narrow field
instruments LECS (Parmar \etal 1997) and MECS (Boella \etal 1997) were
extracted within a circular region with a radius of 8$'$ centered at the source
position. The background for these two instruments were obtained
using blank sky observations. HPGSPC (Manzo \etal 1997) and PDS
(Frontera \etal 1997) data were
extracted using SAXDAS (hpproducts V3.0.0) and XAS respectively.
All our spectral analysis were carried out in the energy range
suggested by the BeppoSAX cookbook\footnote{http://heasarc.gsfc.nasa.gov/docs/sax/abc/saxabc/saxabc.html}: 0.12-4 keV for the LECS,
1.65-10.5 keV for MECS, 8-20 keV for the HPGSPC, 15-220 keV for PDS,
and a 2\% systematic error were added
during the fitting to account for the uncertainties in the calibration.
The public software package $XSPEC$ 11.0.1 was employed in our analysis.

As usual, MCD plus a power-law and
a Gaussian line model was used first to fit the data, which can describe the
observations very well (refer to Figure~\ref{spectralModel}(b)).
The best fit parameters are similar to the previous results reported by
Campana \etal (2002).

Our two table models plus a broad Gaussian line were then applied
to the data, and these two models provide reasonably good fit to all
the BeppoSAX observations (Figure~\ref{spectralModel}(c)-
\ref{spectralModel}(d)) and the best fit parameters are reported
in Table~\ref{modelParameters}. Since we mainly focus our efforts
on the continuum, the line parameters are not listed.

\begin{figure}
  \plotone{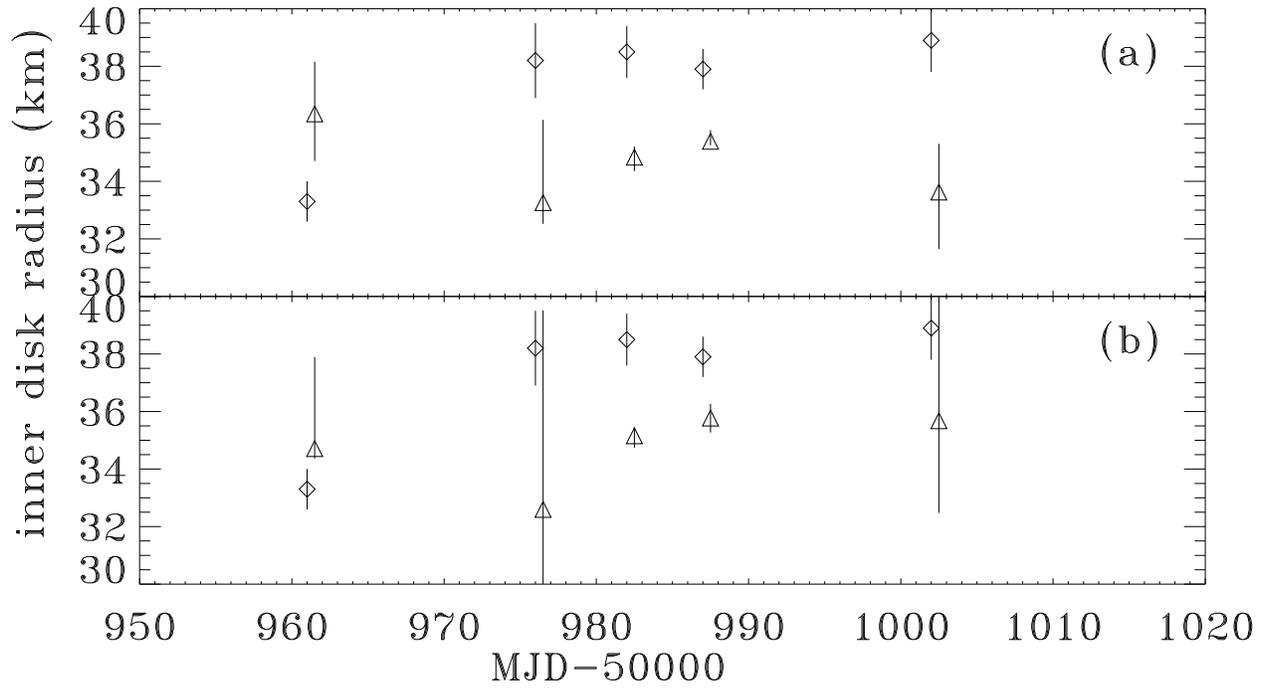} \caption{The inner radius of the accretion disk
    inferred from our table models ({\sl triangle symbols}) for a spherical
    corona system (a) and for a disk-like corona system (b) and those
    derived from spectral-fitting with MCD+PL model ({\sl diamond symbols};
    Campana \etal 2002). In order to
    show the comparison clearly, we shift the observation date slightly in the
    plot.} \label{radius}
\end{figure}

According to the fitting results, the column density nearly
remains constant with a mean central value $1.33 \times 10^{22}$
cm$^{-2}$, and $T_{in}$ is around 0.72 keV, which agree with the
previous report \citep{cam02}. $T_c$ decreases during the BeppoSAX
observations. In the first two observations, the upper limit of
$T_c$ is even beyond the parameter range (current range is 5 keV
to 200 keV) in our table models. In the last three observations,
the spectra are relatively soft and $T_c$ can be constrained
tightly. If the system has a spherical corona, $R_c$ may vary from
several tens to one hundred times of the gravitational radii. The
optical depth $\tau$ is very small ($<0.5$) for all the five
observations, and the $\theta$ values are with a large
uncertainty. The normalization parameter, $K_{norm}$, which is
related to the inner disk radius, also varies dramatically in the
five observations. However, this apparent variation is caused by
the uncertainty of $\theta$ (see discussion below). In order to
further check the variation, we re-normalize $K_{norm}$ to the
value at {\sl zero} inclination angle, and plot the inner disk
radius ($R_{in} = \sqrt{K_{norm} \cos\theta}$ km) for different
observations (Figure~\ref{radius}). For comparison, we also plot
the results from the MCD+PL model \citep{cam02} in the figure.
Even though with large uncertainties, $R_{in}$ derived from our
models are consistent with being constant,
 within the error bars (90\%)
and the ``sharp'' increase
of $R_{in}$ derived from MCD+PL model does not show up in our results.

\section{Discussion}
In this work, we developed a Monte-Carlo code to simulate the
scattering process between the seed photons from an accretion disk
and the electrons in the corona above the disk. Based on the
simulation results, we build two simple table models for the
spherical corona system and for the disk-like corona system, which
can be used in the spectral fitting to directly derive the corona
properties (like its size, opacity, and temperature) as well as
the disk parameters (disk temperature, inclination and inner disk
radius).

The reason that the size of the spherical corona can be determined from the
X-ray spectral fitting is that in our models the seed photons for
the inverse Compton scattering come from the disk, on which the
temperature is a function of the distance from the central black
hole. The relatively small size of the corona of XTE~J2012+381
obtained in this work indicates that the hard photons are mainly
generated within a very small region near the central black hole.

The opacity of the corona can be inferred from the ratio between
the flux of the soft component and that of the hard component,
whereas the temperature of the corona can be determined by the
high energy turn over in the spectrum. However, for XTE~J2012+381,
the $\tau$ we obtained is rather small, indicating only a small
portion of the disk photons were up-scattered to the high
energies. When the source was in its hard spectral state (e.g.,
the first observation), we did not see an obvious turn-over in the
spectrum, which may suggest that the pure thermal corona is not a
good approximation in this case and should be replaced by the
thermal-powerlaw hybrid corona, a more realistic one proposed by
\citet{cop99}. When the source was in its soft spectral state (the
last two observations), we obtained a low corona temperature
$T_c$, which suggests that a large uncertainty may exist in the
inferred corona parameters, because the disk emission could be
mixed with the corona emission and any slight change of the disk
parameter (e.g., inclination angle) could result in apparently
significant change of the corona parameters. Even though these
changes will also cause some changes in the $K_{norm}$ parameters
of our table model, our conclusion marks are based on the
re-normalized $K_{norm}$ (to {\sl zero} degree for example) which
can be hardly impacted.

The reason that the inclination angle can in principle be inferred
is that the observed flux from the disk depends strongly upon the
inclination angle (through a cosine factor), whereas the hard
X-ray flux resulting from Compton scattering is almost isotropic
for the case of spherical corona. For the case of disk-like
corona, the hard X-ray flux also depends upon the disk inclination
angle. However, the inclination angle, which is also related to
the ratio between the soft component flux and the hard component
flux, is always coupled with the optical depth of the corona.
Thus, inferring the disk inclination angle by spectral fitting may
have some ambiguities.

Determining the value of the inner disk radius is important for
understanding the physics of the accretion disk and the black hole
angular momentum (Zhang \etal 1997). The normalization parameter
inferred with our table model, which is proportional to the inner
disk radius squared, is nearly constant in the BeppoSAX
observations, and the previously reported sharp increase of the
disk radius when the source transits from the hard state to the
soft state did not happen. This is because the corona, though
optically thin in most cases, scatters some of the photons emitted
from the disk and makes the observed soft component different
significantly from the original disk emission. Using the simple
MCD+PL model and without considering the origin of the PL like
component will certainly under-estimate the real flux of the
blackbody-like component, especially when the source is in the
hard state. The reason that we can estimate the real flux of the
accretion disk is that the number of the photons emitted from the
disk is known in our Monte-Carlo simulations.

In our simulation process, we did not include the reflection
mechanism to produce a reflection component in the emergent
spectrum, but simply ignored a sampled photon when it collides
with the accretion disk. Even though the ignored photons will not
be reflected in the output spectrum, their corresponding seed
photons are still counted as part of the original disk emission. A
disk reflected spectrum is potentially important for probing the
disk ionization parameter, the geometry, the inclination angle of
the accreting system. When the disk reflected component is clearly
visible in a spectrum (e.g., in Cyg~X-1; Salvo \etal 2001), the
Comptonization model developed in this work plus a reflection
model can be utilized to account for the observed total spectrum.
As long as the combined model can fit the observed spectrum
satisfactorily, the original disk flux could still be
reconstructed from the normalization of the table model.

Please also be noted that in this work we did not consider the
vertical structure of the accretion disk (e.g.,  Shimura \&
Takahara 1995) and the gravitational potential is still assumed to
be the Newtonian potential, therefore, the derived inner disk
radius in this work might be substantially different from the
``real'' value. In fact, because of the high temperature ($\sim 1
keV$) around the inner disk region, the Compton scattering process
may dominate over the free-free process in this region, and the
local emergent spectrum might be approximated as a diluted
blackbody spectrum rather than a real black body spectrum. The
spectral difference due to this effect can be corrected by the
spectral hardening factor, which is $\sim$ 1.7 for an accreting
system at about 10\% accretion rate \citep{shi95}. The correction
for the real potential in the vicinity of the black hole (general
relativity effects) could be done by introducing three factors
$\eta$, $f_{GR}$, $g_{GR}$. The $\eta$ accounts for the difference
between the apparent and the intrinsic radii of the peak
temperature, and the $f_{GR}$ is due to temperature change caused
by the gravitational red shift and Doppler shift, whereas the
$g_{GR}$ is related to the integrated flux change caused by the
gravitational focusing and boosting. The value of these factors,
depending on the system inclination and black hole spin, have been
discussed in several papers \citep{sha73, cun75, ebi94} and also
have been tabulated by \citet{zha97}, which have been proved to be
reasonably accurate \citep{gie99, gie01}. We should point out that
all these corrections, if performed in this study, will
systematically change the values of the derived inner disk radii
but will hardly change the conclusion that the inner disk radius
is nearly constant (this situation has been proved when applying
the same model and the above corrections to the six ultra-luminous
X-ray sources \citep{wang04}).

In our table model, the normalization parameter is very
sensitive to the inclination angle, which, as mentioned before,
 is coupled together
with the optical depth. In order to reduce the uncertainty of the
value of the normalization parameter in our table model and to
infer the real flux of the accretion disk, an accurate value of
the inclination angle is needed and the optical observations may
give useful help on this issue.

\acknowledgments
We thank the anonymous referee for critical, yet insightful comments on our
manuscript, which made us to improve the paper substantially. Mr. Yongzhong Chen is acknowledged for his initial work
on this project during his visit to UAH in 1999-2000. We want to give our
special thanks to Dr. S. Campana for kindly giving us the data of XTE
J2012+381 observed with BeppoSAX. We also thank Drs. Lev Titarchuk and Wei Cui
for interesting discussions and Dr. Alan Harmon for useful comments.
This work was
supported in part by NASA Marshall Space
Flight Center under contract NCC8-200 and by NASA Long Term Space
Astrophysics Program under grants NAG5-7927 and NAG5-8523.

\appendix

\section{Normalization parameter in our table models}
Starting from a multi-color disk black body, assuming the inner disk
temperature is $T_{in}$ and the inner disk radius is $R_{in}$, then
the total number of photons released by the disk per unit time will be
(by integrating Plank function from $R_{in}$ to infinity),
\begin{equation}
\begin{array}{lll}
ph & = & 8\pi a^* T_{in}^3r_{in}^2
\end{array}
\label{diskPhotons}
\end{equation}
where, $a^*=\frac{2\pi}{c^2}(\frac{K}{h})^3 \int^\infty_0 \frac{x^2dx}{e^x -1}$
Thus, the observed photon flux at distance $D$ with inclination angle $\theta$ will be,
\begin{equation}
\begin{array}{lll}
F_{ph}(\theta)ds &=& ph(\theta)d\Omega \\
        & =& 8\cos(\theta) a^* T_{in}^3r_{in}^2 \frac{ds}{D^2} \\
        & =& 8\cos(\theta) a^* T_{in}^3 \left( \frac{r_{in}}{D} \right)^2 ds. \\
\end{array}
\label{observedPhotons}
\end{equation}
If considering the double sides of the accretion disk, there should be
another factor 2 in equation~(\ref{diskPhotons}) and (\ref{observedPhotons}).
However, in our calculation,
we only consider one side and it is self-consistent within our table model.

In our simulation, for each configuration of the parameters, 500,000 photons
are collected. Because the black hole (BH) absorption and the collision
between the photons and the accretion disk (we ignore the disk reflection),
the seed photon number $N_{seed}\ge 500,000$. In each simulation, six spectra
will be produced by collecting the output photons in different directions
from $0^0$ to $35^0$, $35^0$ to $45^0$, $45^0$ to $55^0$, $55^0$ to $65^0$,
$65^0$ to $75^0$, $75^0$ to $85^0$ degrees, corresponding to the inclination
angles $22.7$, $40.1$, $49.9$, $59.8$, $69.6$, $79.2$ degrees respectively
(by considering the solid angle and disk projection).
To make a connection between the seed photon number in our simulation and the
real disk flux, each spectrum had been scaled down
by a factor $f_{scale}$, which is determined by,
\begin{equation}
 f_{scale} = N_{seed} / ph\\
\end{equation}
To get the spectrum observed at different inclination angles, each
angle-dependent spectrum has been scaled down by different area factors
$\Delta s_{\theta}$ to get the average flux at that inclination angle,
which is determined by,
\begin{equation} \begin{array}{lll}
\Delta s_{\theta} &=& \int^{\theta _j}_{\theta _i} D^2 \sin(\theta)d \theta \int^{2\pi}_0 d \phi \\
 &=& 2\pi D^2 \left[ \cos(\theta _i) - \cos(\theta_j) \right]
\end{array}
\end{equation}
If we take $r_{in}=1$ km, $D = 10$ kpc in equation (\ref{observedPhotons}) (in
order to compare with $diskbb$ in $XSPEC$), the relation between real
$r_{in}$ and normalization of the table model is,
\begin{equation}
K_{norm} = \left( \frac{r_{in}/km}{D/10kpc} \right)^2
\end{equation}

\clearpage


\clearpage

\clearpage

\clearpage

\clearpage

\clearpage
\begin{table}
\begin{center}
\caption{Results from spectral fits with our table models.}
\label{modelParameters}
\setlength{\evensidemargin}{1in}
\footnotesize
\begin{tabular}{cccccccccc}
\hline
\hline
Date & $N_H$ & $T_{in}$ & $T_c$& $R_c$ &  & $\theta$ & $K_{norm}\tablenotemark{a}$ &  $\chi^2$ \\
(MJD) & (${\rm 10^{22}cm^{-2}}$) & (keV) & (keV) & (${\rm R_g}$) & $\tau$ & ($^0$)& (${\rm km^2}$) & (/dof) \\
\hline
\multicolumn{9}{l}{Model A: spherical corona}\\
\hline
$50961$  & $1.37^{-0.01}_{+0.02}$ &  $0.729^{-0.001}_{+0.003}$ &  $200^{-9}_{+**}$      &  $20^{-4}_{+4}$       &  $0.227^{-0.028}_{+0.030}$ &  $25^{-**}_{+21}$ &  $1452^{-128}_{+148}$ &  $292(/242)$ \\
$50976$  & $1.33^{-0.03}_{+0.03}$ &  $0.707^{-0.004}_{+0.005}$ &  $200^{-38}_{+**}$     &  $10^{-**}_{+16}$     &  $0.111^{-0.111}_{+0.016}$ &  $23^{-**}_{+40}$ &  $1202^{-53}_{+217}$ &  $207(/170)$  \\
$50982$  & $1.33^{-0.02}_{+0.03}$ &  $0.713^{-0.003}_{+0.004}$ &  $100^{-19}_{+10}$     &  $49^{-13}_{+18}$     &  $0.070^{-0.006}_{+0.004}$ &  $59^{-18}_{+5}$   &  $2357^{-65}_{+50}$  &  $179(/195)$ \\
$50987$  & $1.33^{-0.01}_{+0.03}$ &  $0.730^{-0.002}_{+0.006}$ &  $5.8^{-**}_{+ 0.8}$   &  $99^{-15}_{+10}$     &  $0.098^{-0.004}_{+0.019}$ &  $59(fix)$        &  $2432^{-19}_{+54}$  &  $197(/148)$ \\
$51002$  & $1.31^{-0.01}_{+0.03}$ &  $0.709^{-0.003}_{+0.003}$ &  $5.5^{- 0.5}_{+ 1.0}$ &  $99^{-20}_{+20}$     &  $0.097^{-0.017}_{+0.017}$ &  $29^{-**}_{+46}$ &  $1293^{-148}_{+132}$ &  $175(/149)$ \\
\hline
\multicolumn{9}{l}{Model B: disk-like corona}\\
\hline
$50961$  & $1.35^{-0.02}_{+0.01}$ &  $0.722^{-0.003}_{+0.007}$ &  $200^{-22}_{+**}$     &  --                   &  $0.116^{-0.007}_{+0.009}$ &  $18^{-**}_{+16}$ &  $1260^{-25}_{+241}$  &  $278(/243)$ \\
$50976$  & $1.31^{-0.03}_{+0.02}$ &  $0.708^{-0.003}_{+0.002}$ &  $200^{-40}_{+**}$     &  --                   &  $0.015^{-0.003}_{+0.017}$ &  $69^{-**}_{+4}$  &  $2957^{-1231}_{+1388}$   &  $208(/171)$ \\
$50982$  & $1.33^{-0.02}_{+0.02}$ &  $0.713^{-0.002}_{+0.001}$ &  $100^{-13}_{+9}$      &  --                   &  $0.050^{-0.005}_{+0.007}$ &  $18^{-**}_{+28}$ &  $1292^{-30}_{+23}$  &  $182(/196)$ \\
$50987$  & $1.34^{-0.01}_{+0.01}$ &  $0.720^{-0.002}_{+0.001}$ &  $10.5^{-0.5}_{+1.2}$  &  --                   &  $0.222^{-0.013}_{+0.021}$ &  $53^{-12}_{+16}$ &  $2126^{-59}_{+59}$   &  $193(/149)$  \\
$51002$  & $1.32^{-0.02}_{+0.01}$ &  $0.701^{-0.004}_{+0.001}$ &  $13.0^{-0.7}_{+1.5}$  &  --                   &  $0.281^{-0.061}_{+0.085}$ &  $37^{-**}_{+26}$ &  $1553^{-266}_{+924}$   &  $161(/150)$ \\
\hline
\hline
\end{tabular}
\tablenotetext{**}{the limit is not reachable.}
\tablenotetext{a}{Normalization of the table model,
  $K_{norm}=((R_{in}/{\rm km})/(D/10 {\rm kpc}))^2$.}
\end{center}
\end{table}

\end{document}